\documentclass[jocse]{jocseart}

\usepackage{booktabs} 
\usepackage{subfig}

\setcopyright{jocsecopyright}

\jocseDOI{10.22369/issn.2153-4136/x/x/x }

\begin{document}
\title{Deep Learning by Doing: The NVIDIA Deep Learning Institute \\
and University Ambassador Program}

\author{Xi Chen}
\orcid{0000-0001-7094-6748}
\affiliation{%
  \institution{University of Kentucky}
  \streetaddress{}
  \city{Lexington}
  \state{Kentucky}
  \postcode{40508}
}
\email{billchenxi@gmail.com}

\author{Gregory S. Gutmann}
\affiliation{%
  \institution{Tokyo Institute of Technology}
  \streetaddress{}
  \city{Tokyo}
  \state{Japan}
}
\email{gutmann@c.titech.ac.jp}

\author{Joe Bungo}
\affiliation{%
  \institution{Deep Learning Institute, NVIDIA Corporation}
  \streetaddress{}
  \city{Austin}
  \state{Texas}
}
\email{jbungo@nvidia.com}


\begin{abstract}
Over the past two decades, High-Performance Computing (HPC) communities have developed many models for delivering education aiming to help students understand and harness the power of parallel and distributed computing. Most of these courses either lack a hands-on component or heavily focus on theoretical characterization behind complex algorithms. To bridge the gap between application and scientific theory, NVIDIA Deep Learning Institute (DLI) (\url{nvidia.com/dli}) has designed an on-line education and training platform  that helps students, developers, and engineers solve real-world problems in a wide range of domains using deep learning and accelerated computing. DLI's accelerated computing course content starts with the fundamentals of accelerating applications with CUDA and OpenACC in addition to other courses in training and deploying neural networks for deep learning. Advanced and domain-specific courses in deep learning are also available. The online platform enables students to use the latest AI frameworks, SDKs, and GPU-accelerated technologies on fully-configured GPU servers in the cloud so the focus is more on learning and less on environment setup. Students are offered project-based assessment and certification at the end of some courses. To support academics and university researchers teaching accelerated computing and deep learning, the DLI University Ambassador Program enables educators to teach free DLI courses to university students, faculty, and researchers.
\end{abstract}


\maketitle

\section{Introduction}

With an increasing emphasis on using computing as the new scientific experimentation method, computer science has become an interdisciplinary academic area. Deep Learning (DL), an emerging and powerful tool for machine learning, has gained attention in recent years due to its potential to reshape the future of computational research. Many ground-breaking results rely on DL, such as image classification\cite{Szegedy:2015tb}, Atari game bots\cite{Silver:2016hl}, and medical diagnosis\cite{Ravi:dw}, and have achieved superhuman levels of performance. One reason for this recent quantum leap in research has its roots in the growing prevalence of High-Performance Computing (HPC).  

	HPC provides the parallel and distributed computing power that allows deep learning to excel. NVDIA CUDA-powered GPUs have several times enabled the title of the fastest supercomputer in the world since 2010\cite{Cook:2013cq}. Since then, the new setup for most HPC research has been parallel systems incorporating CPUs with GPUs\cite{Nickolls:2008hn}. 

	Despite decades-long efforts in HPC education, HPC and DL are still understood and used by only a small portion of the scientific and engineering community\cite{Mullen:2017ih}. There two main problems contributing to the low rates of adoption: 1. the material requires a certain amount of computational and statistical literacy; and 2. HPC programming environments are difficult to set up due to the variety of hardware and OS system types. Overcoming these two concerns requires curating learning material for a broad range of audiences that includes practical cases, as well as abstracting the system implementation to simplify the use of HPC resources.  
    
	Combing current online learning andragogy with a cloud computing platform consisting of a VM and Docker containers, we illustrate a new educational platform designed by the NVIDIA Deep Learning Institute as showing in Figure 1. This platform provides hands-on experience with DL and facilitates practical DL and HPC education.

\begin{figure}
\includegraphics[width=0.45\textwidth]{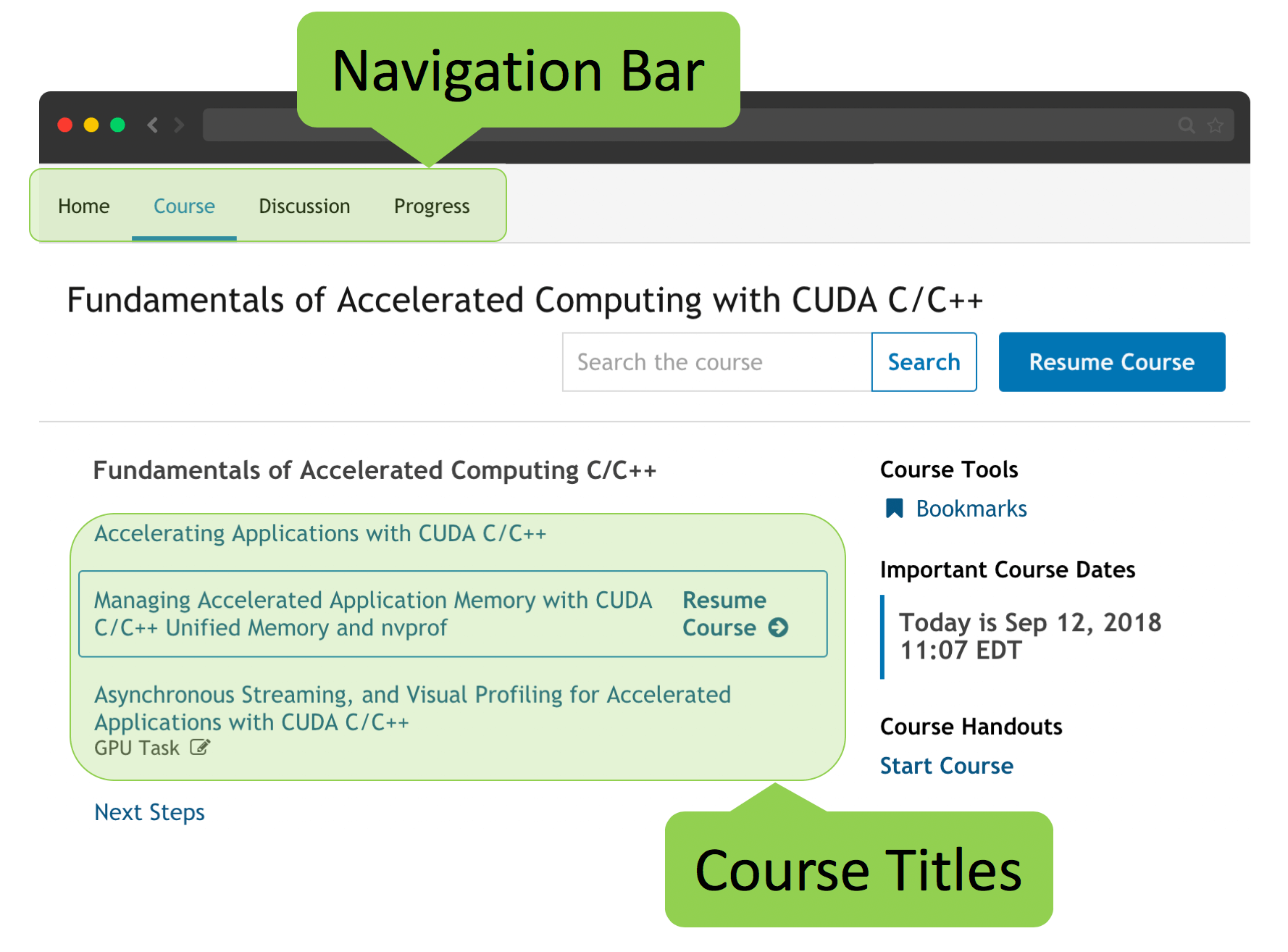}
\caption{CUDA C/C++ course on DLI online platform. The Navigation Bar allows students navigate across different sections of the platform. Courses that contain multiple sub-sections will list all course titles as url links to navigate to the course content.}
\end{figure}

\subsection{High Performance and GPU Accelerated Computing Education}
	High Performance Computing and GPU acceleration used to be accessible only to scientists and engineers to fulfill their desire to better model realistic physical systems. As computing hardware becomes cheaper and its performance improves, these resources are no longer found only in mega research centers. Major tech companies such as Microsoft, Google, and Amazon now provide low-cost HPC instances in the cloud, and major academic programs have gradually embraced the convenience and efficiency of HPC because of it. Despite these efforts there is still a large unmet need for new, powerful, HPC-enabled tools and curricula for education.

	To extend exposure and build a larger base within the HPC research community, many government agencies have initiated programs aimed to develop awareness at all stages of the education pipeline. These include the Education, Outreach and Training (EOT) program from the National Science Foundation (NSF), the Advanced Scientific Computing Initiative (ASCI) funding program, and the High-Performance Computing Modernization Program (HPCMP) from the U.S. Department of Energy (DOE), among others. Such initiatives have a huge effect in developing awareness of HPC at all stages of the education system. 
	
    While waiting for these efforts to pay off, and in response to the immediate need for HPC, hundreds of modules and online training sessions have been created for users from the world's supercomputing centers. While these approaches are widespread and effective, they cannot be scaled to reach a larger audience as they are centered around a set of one-to-one tutorials and are only available for certain infrastructures. In addition, due to the high demand of research needs, only limited HPC resources from these computing centers are allocated  for education and experimentation by the beginning learners and students.

\subsection{Abstractions through CUDA}
CUDA, in essence, is a minimal extension of the C and C++ programming language. As a scalable parallel programming platform, it allows sophisticated HPC programing to be expressed in an easily understood abstraction. After its release in 2007, CUDA rapidly evolved into a popular Application Programming Interface (API) model that facilitates a wide range of research and applications\cite{Luebke:ih}. 
	
    CUDA provides three key abstractions: a hierarchy of thread groups, shared memories, and barrier synchronization. The CUDA paradigm allows programmers to partition a "problem into coarse subproblems that can be solved independently in parallel, and then into finer pieces that can be solved cooperatively in parallel"\cite{Nickolls:2008hn} while hiding memory management and thread synchronization behind the scenes. In addition, NVIDIA provides a CUDA profiler to help programmers further understand and debug each parallel process to achieve the best possible performance (Figure 7).

\subsection{NVIDIA's Deep Learning Institute for HPC and DL Education}
NVIDIA's Deep Learning Institute (DLI) aims to lower the barrier of entry for HPC education and develop novel ways of enabling scientists and engineers in their own research through the use of HPC resources [\url{www.nvidia.com/dli}]. Based on the experience from previous instructor-led courses via DLI's University Ambassador Program, cloud-based computing solutions for training and education reduces the learning curve and allows students to gain hands-on experience with parallel computing systems without getting bogged down setting up programming environments. Participants can earn certification to prove subject matter competency and support professional career growth. Certification is offered for select online courses and instructor-led workshops. [\url{www.nvidia.com/dli}] All students need to participate in the hands-on training is a computer with a modern browser installed and a reliable internet connection. 
\begin{figure}%
    \centering
    \subfloat[To start a course, click the "Start" button.]{{\includegraphics[width=3.2cm]{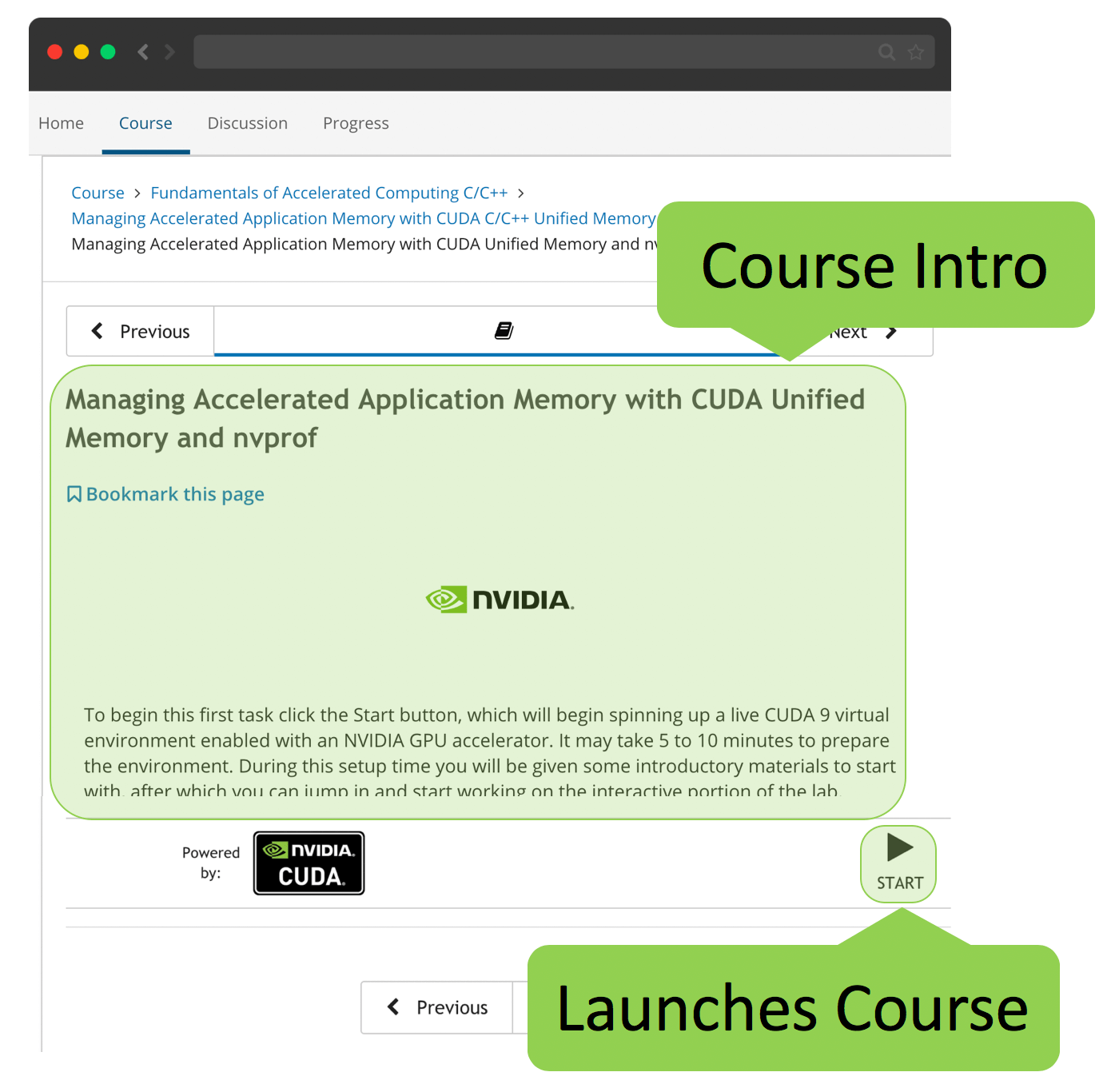} }}%
    \qquad
    \subfloat[How to access Jupyter Notebook. To stop GPU server and end a course, click "Stop Task."]{{\includegraphics[width=3cm]{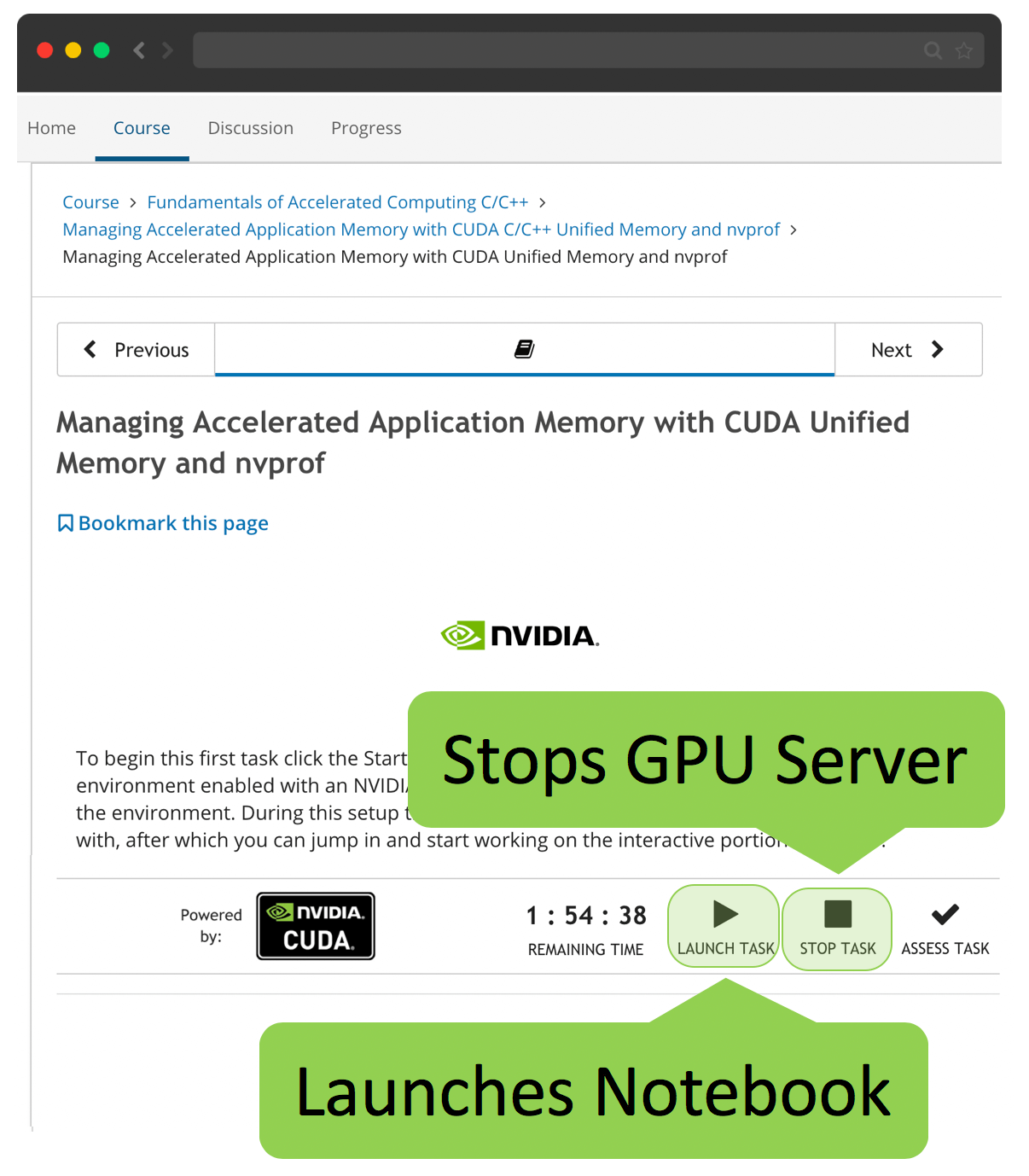} }}%
    \caption{Individual course landing page.}%
    \label{Figure 2}%
\end{figure}
	DLI is currently using GPU virtual instances in combination with Jupyter Notebooks to implement a hands-on, project-based experience in DL and accelerated computing. Participants can easily click a "Launch Task" or "Stop Task" button to begin and end their learning tasks as part of a full course, as showing in Figure 2. If a student has issues or questions about the course material at any point, they can participate in a MOOC-style discussion session to post their questions or search for solutions, as showing in Figure 3. For programming questions beyond the scope of the course, NVIDIA provides online forums and documentation that allow students to find answers to more complex questions.
    
\begin{figure}%
    \centering
    \subfloat[Users can post a topic (A), search existing topics (B), and   browse all topics (C). The instruction of how to use Discussion (D).]{{\includegraphics[width=3.5cm]{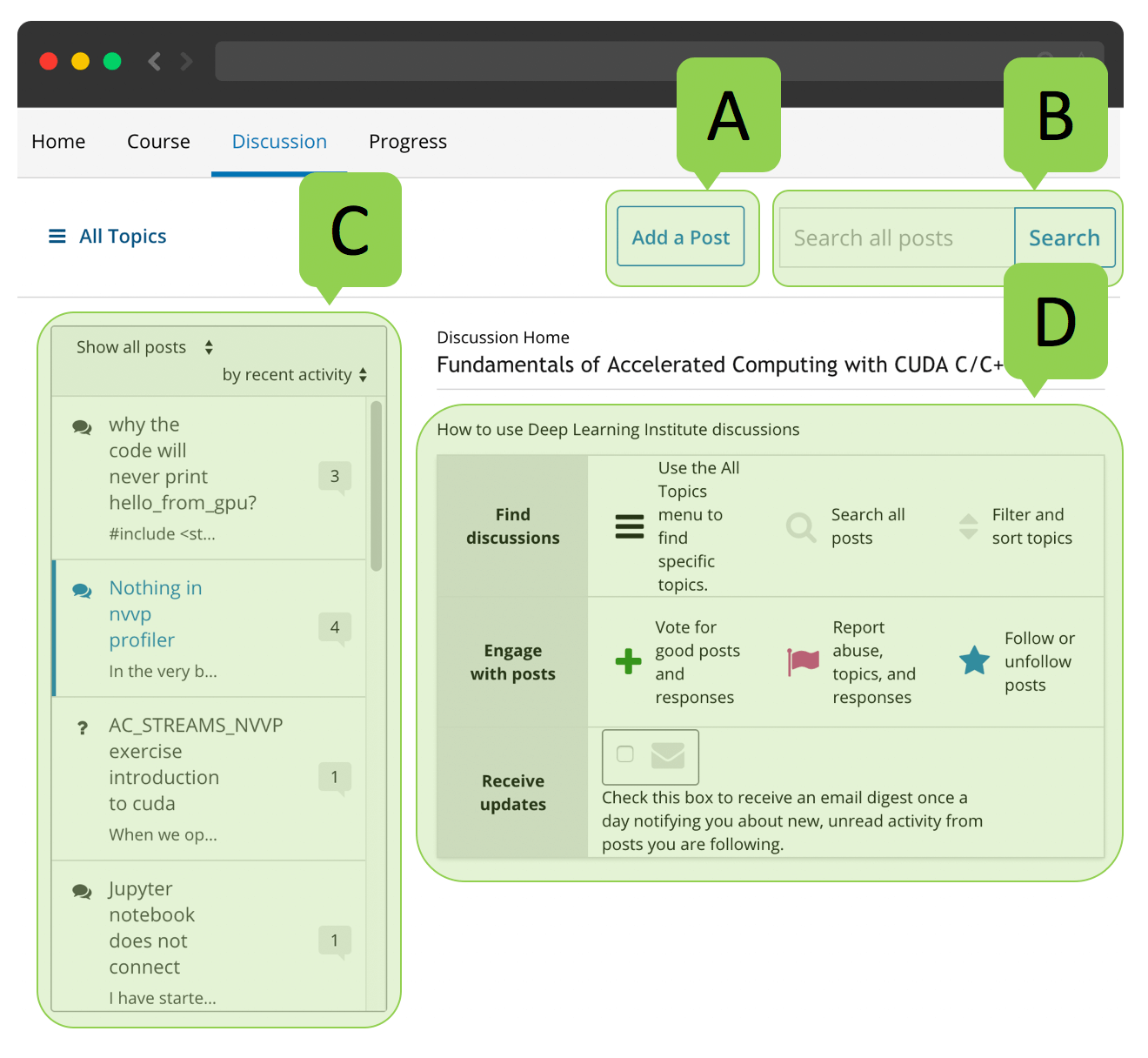} }}%
    \qquad
    \subfloat[Individual topic layout: (A) Title and content of the topic, and (B) responses from members or course staff.]{{\includegraphics[width=3.5cm]{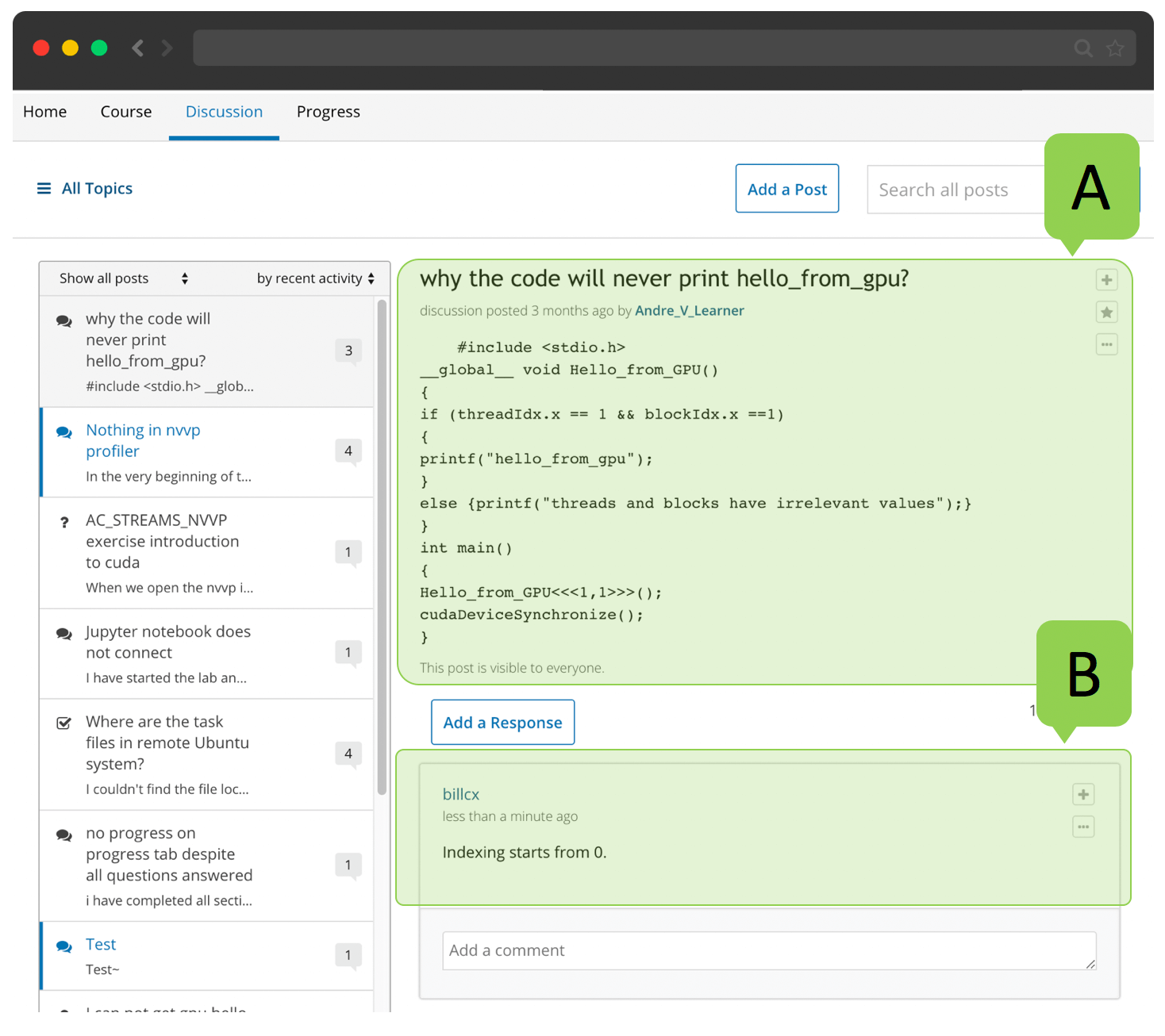} }}%
    \caption{Discussion panel layout.}%
    \label{Figure 3}%
\end{figure}

DLI offers five full-day, hands-on "Fundamentals" courses for those who may be new to accelerated computing or deep learning. Two of DLI's most popular courses with project-based certification are Fundamentals of Deep Learning for Computer Vision and Fundamentals of Accelerated Computing with CUDA C/C++. To prepare learners with more practical knowledge, DLI also provides application-specific content in the following disciplines:

\begin{itemize}
\item Deep Learning for Autonomous Vehicles
\item Deep Learning for Healthcare
\item Deep Learning for Digital Content Creation
\item Deep Learning for Finance
\end{itemize}

In this article, we present a hands-on, online course provided by DLI: Fundamentals of Accelerated Computing with CUDA C/C++ (Figure 4). This course provides a path that allows learners with little or no knowledge of HPC GPU acceleration to harness the power of parallel computing and possibly achieve an official CUDA certification based on project-based assessments.

	The rest of the article is organized as follows: In Section 2, we present the technologies used in the design of the course. Section 3 highlights the University Ambassador Program, and Section 4 demonstrates the results of teaching events across the globe. We conclude with a summary of the work.
    
\begin{figure}
\includegraphics[width=0.45\textwidth]{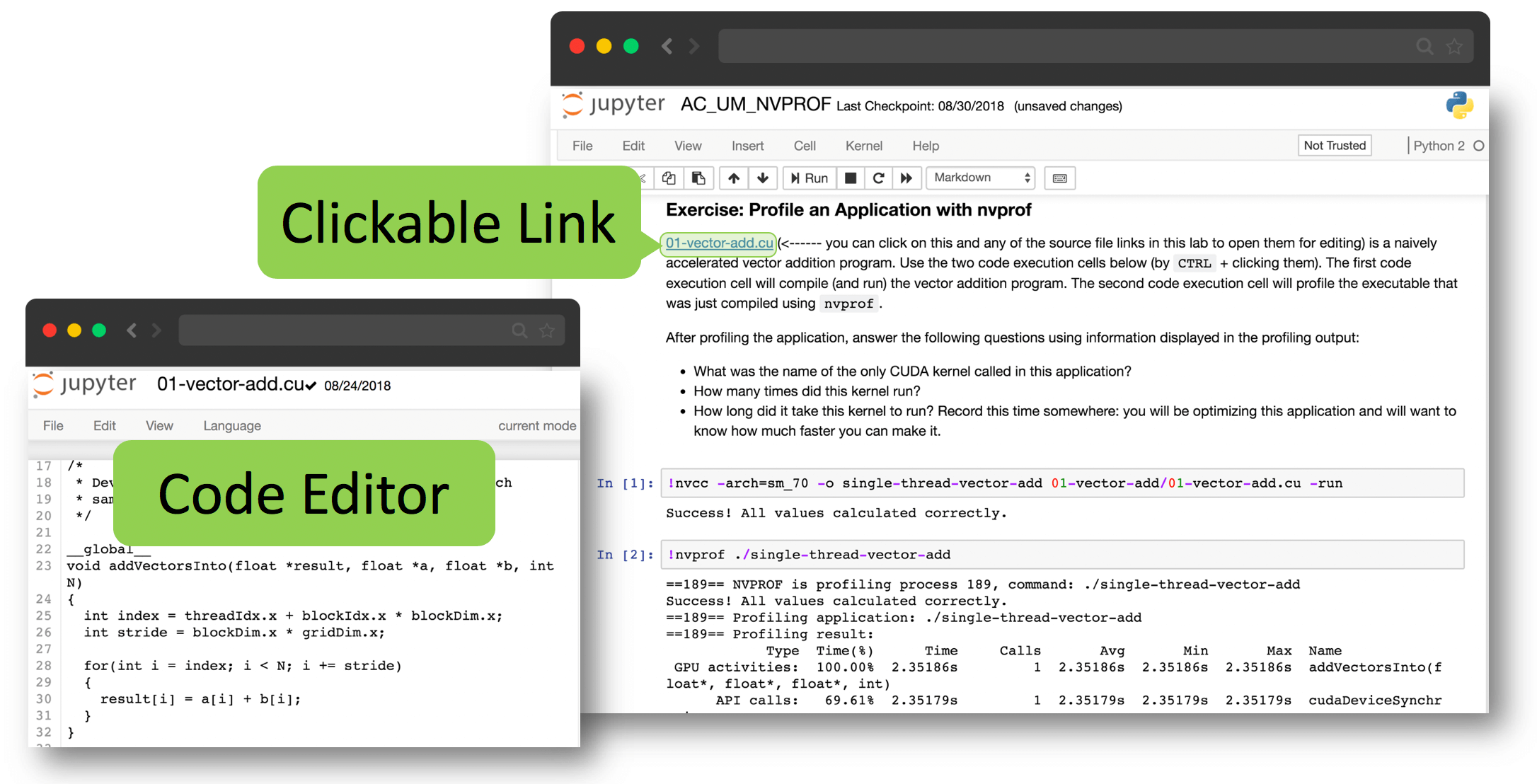}
\caption{Inside the DLI CUDA C/C++ course - Managing Accelerated Application Memory with CUDA C/C++ Unified Memory and nvprof. Course content including links that allow students open browser-based code editors.}
\end{figure}

\section{Methods}
When developing courseware, DLI combines the primary design elements of HPC abstraction and hands-on experience with deep learning. Before DLI courses, the vast majority of learners were using personal computers or relying on traditional batch-processing application programming interfaces (APIs) to manage a HPC environment. Instead of convincing participants to follow the traditional HPC education paradigm, DLI focused on providing access to fully-configured GPU servers in the cloud that abstract complicated environment setup, HPC project-based courses and tasks available to students perpetually after training for students to refer back to course materials at any time. 
	
    Since NVIDIA's 2007 introduction of CUDA, a parallel computing platform API, and the emergence of many open source software libraries extended from CUDA such as TensorFlow, TensorRT, Torch, Caffe, CNTK, etc., access to high-performance computation has given researchers the scientific and parallel computing abstractions they need to utilize HPC resources. Therefore, all DLI courses and labs provide CUDA-powered GPU acceleration. NVIDIA also provides cloud-computing support through the NVIDIA GPU Cloud (NGC): GPU-accelerated containers available on-demand on all major cloud platforms for accelerating deep learning and scientific research. 

\subsection{Reflection-in-action Model}
From the very beginning of DLI's course development, the emphasis has always been on hands-on experience and a journey of reflection-in-action. For participants of all backgrounds, learning by doing provides the ultimate method to master a skill, especially in the case of continuing learning throughout a professional career. For live, instructor-led courses, students receive active coaching and participate in a transparent teaching experience with a instructor which helps students understand how and why they are learning course content in particular ways. A similar experience is  provided by DLI online self-paced courses. By presenting students with real-world problems, feedback and project-based assessment, students develop a much better grasp of the course material. A trial-and-error approach to solve in-lecture tasks provides students with opportunities for self-discovery and self-appropriate learning [Schon 1987] (Figure 5). We believe a hands-on andragogy encourages students to go beyond what the course work provides and makes it more likely that they will later apply what they have learned in their own disciplines. 

\begin{figure}
\includegraphics[width=0.3\textwidth]{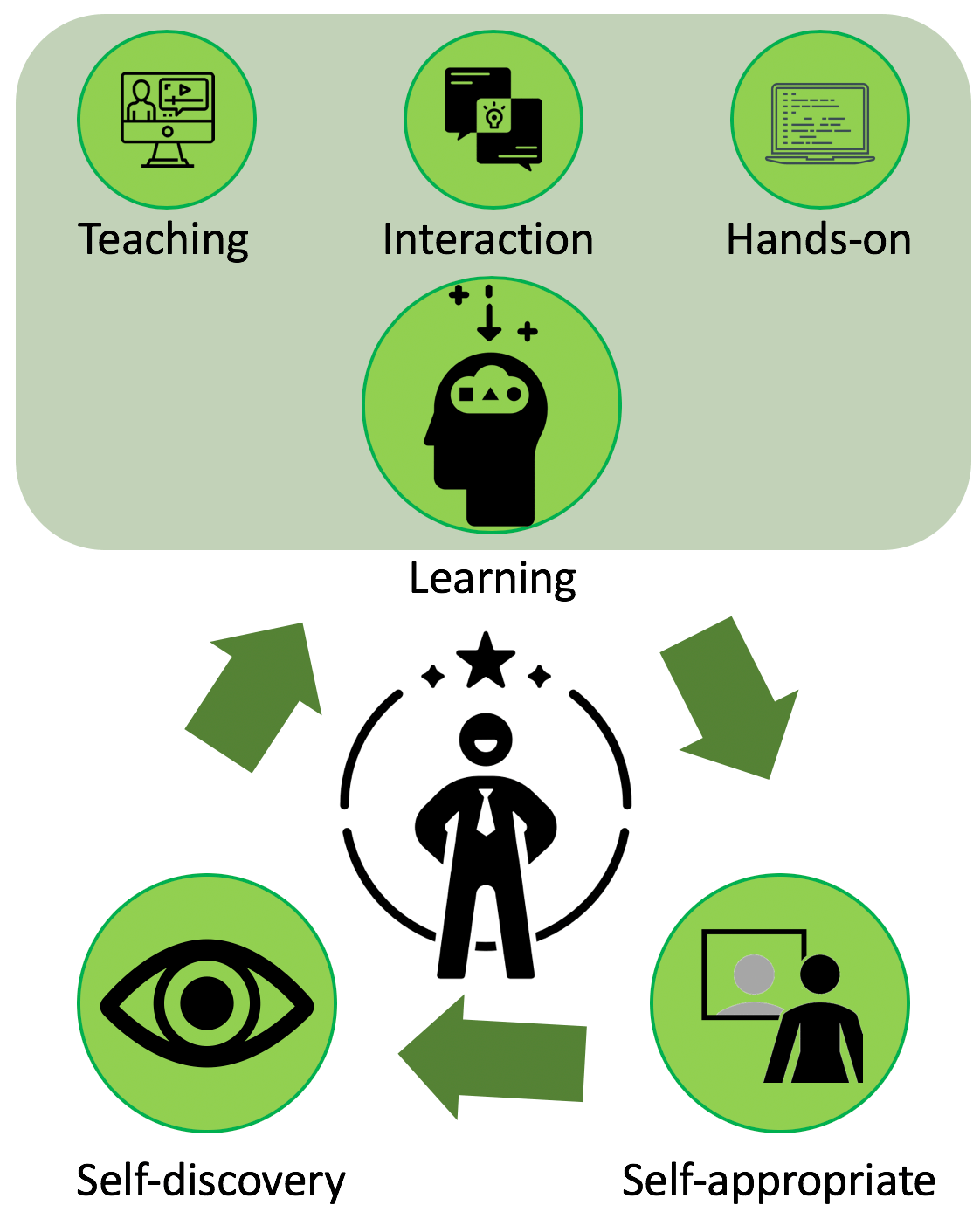}
\caption{By combining teaching and interaction with instructor and hands-on practice, learning promotes self-discovery and self-appropriate. which leads to future application.}
\end{figure}

	If students are enthusiastic and want to immediately apply what they have learned, NVIDIA also provides documentation webpages, FAQs, and expert support to facilitate the efforts of course participants. 

\subsection{OpenEdX MOOC and Cloud Delivery Method}
Massive Open Online Courses (MOOCs) are no longer a new concept in the educational landscape. Open edX, a non-profit MOOC platform, emerged in 2012 [Porter et al. n.d.]. The main aim behind it is to provide essential education for a particular topic to anyone, and sometimes at any time,either toward a degree or just to satisfy one's own personal learning desires. One advantage of this approach is the ability to reach many attendees across geographical location and technology barriers. The Open edX platform includes both a Content Management System (CMS) and a Learning Management System (LMS)\cite{Porter:92vs5iLn}. DLI has adopted this platform and delivers courses to help address the pressing educational needs in the HPC and communities.

	Unlike traditional education, MOOC-style platforms  provide an immediate feedback assessment technique (IF-AT) that allows students to receive immediate results that assess their knowledge\cite{Ho:2015ee}. IF-AT encourages students to self-discover and self-explore new knowledge, provides an enjoyable learning experience, and evidently improves students' retention of the course content\cite{Epstein:2017ij}.
    
    To date, DLI have planned and delivered hundreds of instructor-led trainings, created more than 50 courses and labs across the globe, and have fully embraced the MOOC paradigm.

	One obstacle that many HPC and parallel computing beginners face is the environment setup. To learn from a hands-on programming project, students need a clean and functional setup so the debugging process can focus on coding problems instead of HPC environment issues. DLI uses a cloud container solution to allow students to focus more on learning and less on environment set-up. DLI uses fully-configured GPU servers in the cloud to provide this immersive programming environment. Cloud containers are a lightweight technology to virtualize applications in the cloud. They allow elastic and rapid resource pooling to provide a fully functional parallel CPU and GPU programming environment\cite{Bernstein:2014dv}. The average setup time is below 5 minutes for most DLI courses. To further simplify the process for students, DLI uses an automated script to abstract the whole process behind the scenes as showing in Figure 2.
    
\subsection{Interactive Course Interface}
Here we illustrate the DLI interface using the Fundamentals of Accelerated Computing C/C++ course as an example. Students navigate through the course via a series of links. Each course page links to areas within four subgroups: Home (the landing page), Course, Discussion, and Progress (Figure 1). The actual training content is located in the course section and developed into modules using flecture videos and practical/programming sessions to reinforce the key concepts, as shown in Figure 1. Navigating between units and modules is as simple as clicking on the links, completing assessments, and leaving the interactive assessment lab by clicking the start/launch and stop buttons (Figure 2).

	As part of the LMS, a progress page keeps track of all the questions and projects that a student has completed as well as the corresponding assessment results for each module. This allows both students and instructors have an at-a-glance grasp of the students performance to date. The hands-on assignments (Figures 4) were developed using Jupyter notebooks allowing students to receive immediate feedback while working though the content. In the case of the CUDA C/C++ course, the remote NVIDIA Visual Profiler (Figure 6) enabled students student a deep understanding from the GPU visual metrics allowing them to better determine the bottlenecks in a GPU function.

\begin{figure}
\includegraphics[width=0.45\textwidth]{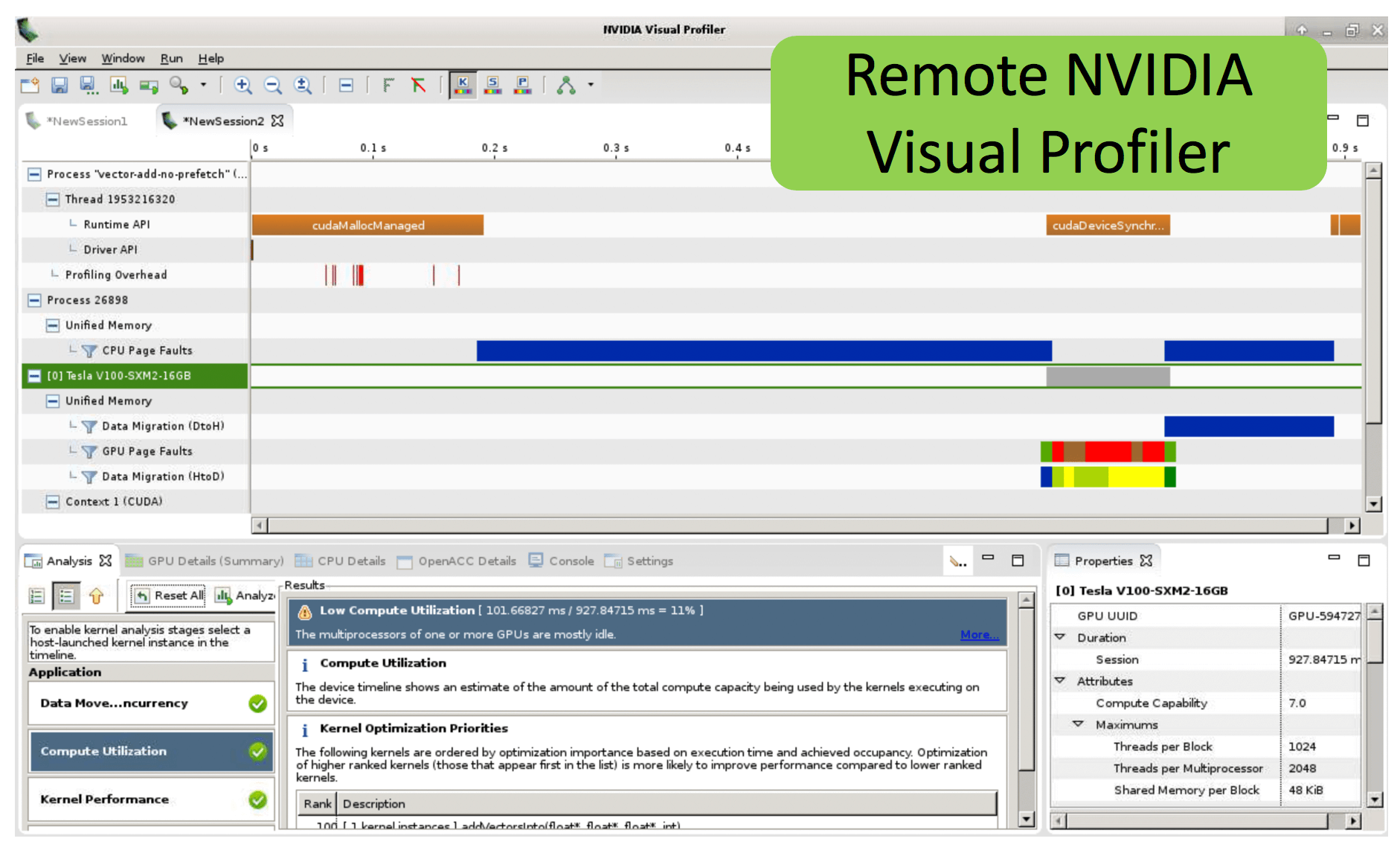}
\caption{Remote Nvidia Visual Profiler captures all of the GPU metrics.}
\end{figure}

\section{DLI University Ambassador Program}

\subsection{DLI Mission}
The mission of the DLI is to help engineers and researchers solve extremely challenging problems using AI and deep learning. To achieve this end, it is necessary to utilize HPC platforms. DLI provides the education essential for using such platforms leveraging massively parallel GPUs. DLI helps "developers, data scientists and engineers to get started in architecting, optimizing, and deploying neural networks to solve real-world problems in diverse industries such as autonomous vehicles, healthcare, robotics, media \& entertainment and game development." [\url{nvidia.com/dli}]

	As stated above, one of the most important elements of the DLI course design is the hands-on experience, which is integrated into both the self-paced and instructor-led versions of the courses. Course materials are currently available across many disciplines ranging from finance to biology, from computer vision to autonomous vehicles, and from game development to accelerated computing [\url{nvidia.com/dli}]. More industry-specific content is coming soon.

\subsection{Bringing DLI to Campus: University Ambassador Program}
DLI recognizes and awards qualified academics as applied deep learning experts. "DLI University Ambassador" is an additional status academics and researchers can achieve on top of a DLI instructor certification. The University Ambassador Program enables educators to teach free instructor-led DLI courses exclusively to university students and staff. This program is free to join and provides a wealth of benefits to academic communities looking to bring AI and deep learning to their campuses. The Ambassador Program offer free DLI instructor certification. provides ready-made, world-class educational content to universities, and offers expense reimbursement for travel and catering expenses for instructor-led workshops. Ambassadors leverage DLI content in their university curriculum courses, campus-wide workshops, and to satisfy workshop and tutorial submissions at academic conferences around the world. For more information about this program visit \url{www.nvidia.com/dli}. 

\subsection{DLI Teaching Kits}
In an effort to extend the HPC community and encourage students to harness the power of DL, DLI offers downloadable Teaching Kits co-developed with well-known academic experts and universities. Each kit includes curriculum materials for a semester-long university course. Complementing DLI's application and hands-on approach, Teaching Kit content integrates more academic theory to satisfy the needs of traditional university coursework. Albeit, in addition to lecture slides, video, and textbook materials, there is at least one accompanying hands-on lab with full source code solutions found in private Git repositories. Many Teaching Kit modules offer multiple labs and solutions. The Teaching Kit program also enables educators to give free access to online, self-paced DLI courses and student certification by way of promotional codes on the MOOC-style platform. To learn more about these resources please visit developer.nvidia.com/teaching-kits. The current Teaching Kits and academic co-authors includes:
\begin{itemize}
\item Machine/Deep Learning (NYU/Yann LeCun)
\item Accelerated/Parallel Computing (UIUC/Wen-Mei Hwu)
\item Robotics (CalPoly)
\end{itemize}

Most Teaching Kits contain:
\begin{itemize}
\item Lecture slides
\item Lecture videos
\item Hands-on labs with solutions
\item Larger coding projects with solutions
\item Quiz/Exam questions with solutions
\item Electronic textbooks
\item DLI online self-paced promotional codes
\item Syllabus with specific DLI online labs interleaved 

\end{itemize}

\subsection{Fundamentals of Accelerated Computing with CUDA C/C++}
In the Fundamentals of Accelerated Computing courses, DLI introduced CUDA parallel computing platform that aims to accelerating computing in terms of impressive performance and ease of use. CUDA supports many popular programming languages such as C, C++, Fortran, Python and MATLAB and expresses parallelism through extensions in the form of basic keywords. CUDA has an ecosystem of highly optimized libraries for DNN, BLAS, graph analytics, FFT, and more, and also ships with powerful command line and visual profilers. Here we present a glimpse of the course to help readers have a better understand of the structure and contents of the course on how to use CUDA to accelerate computing in C/C++.

To help students ease in the parallel programming paradigm, the CUDA course first differentiates the GPU-accelerated vs. CPU-only applications through a series of animation as showing in Figure 7. Then after CUDA syntax and keywords introduction, a hands-on task follows: students need to modify a CPU C/C++ function into a GPU kernel as showing in Figure 8. (Due to the space limitation, we will not further include detailed and exciting course contents.) For every hands-on task, comments in the code will assist student work and the solutions are available in case they get stuck. 

\begin{figure}
\includegraphics[width=0.45\textwidth]{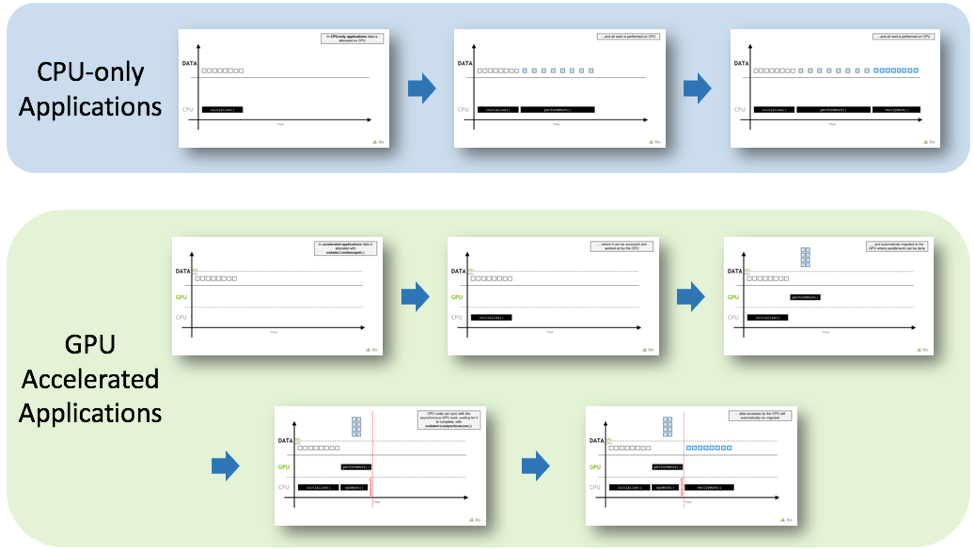}
\caption{Slides on differentiating GPU-accelerated vs. CPU-only applications.}
\end{figure}

\begin{figure}
\includegraphics[width=0.45\textwidth]{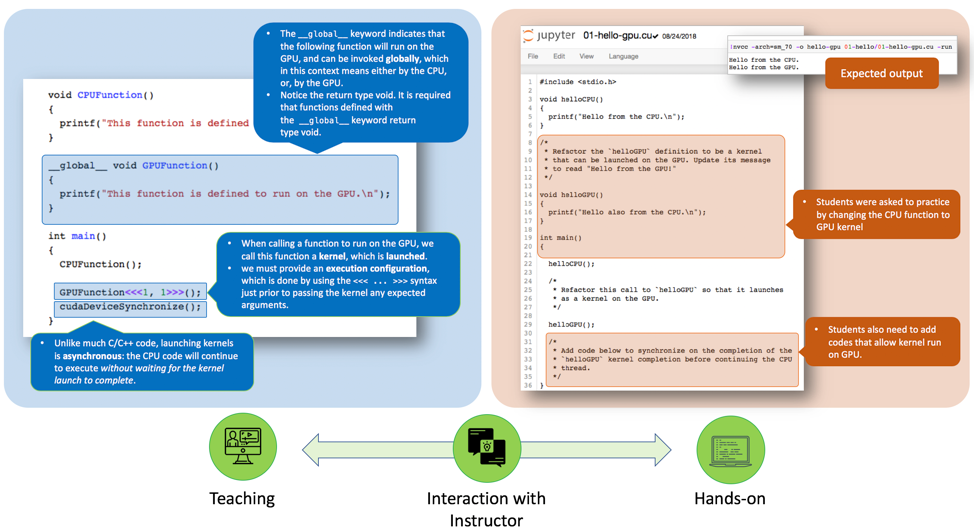}
\caption{Hands-on experience reinforces learning. Through hands-on coding practice, students have chance to learn though doing. It also opens opportunities to interact with instructor.}
\end{figure}

Following the CUDA design cycle: Assess Parallelize, Optimize, Deploy (APOD), in the second and third labs of the course, students will learn how to use command line and visual profilers to further optimize their CUDA code. Along the way, CUDA course will also introduce concepts such as Unified Memory, Streaming Multiprocessors, asynchronous memory prefetching, manual memory allocation and copying, Streams and so on. All in the efforts to help students to have a better grasp of the HPC programming bottle-necks in real-world scenario and to offer solutions on how to toggle them. In Figure 9 we showed that using the command-line and visual profilers one can quickly and qualitatively measure the performance of a application, and to identify opportunities for optimization.

\begin{figure}
\includegraphics[width=0.45\textwidth]{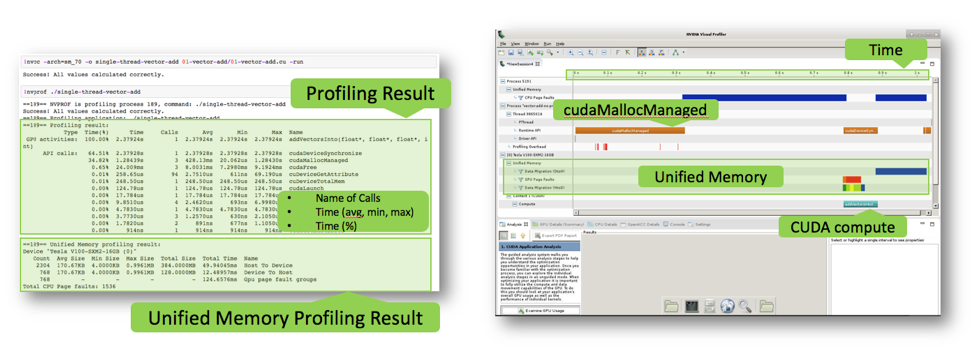}
\caption{Command line and Visual Profilers. In second and third lab, students will learn how to use profiler to measure the performance and identify opportunities for optimization.}
\end{figure}

After students finish the course and pass the assessments, they will receive a certificate of course completion, as shown in Figure 10. Every certification has a unique identification number and online hyperlink, and is tied directly to the student. Students are actively linking their certifications from their resumes and LinkedIn profiles. At the time of this writing, there are no other hands-on, project-based assessment and certification programs in applied deep learning.

\begin{figure}
\includegraphics[width=0.25\textwidth]{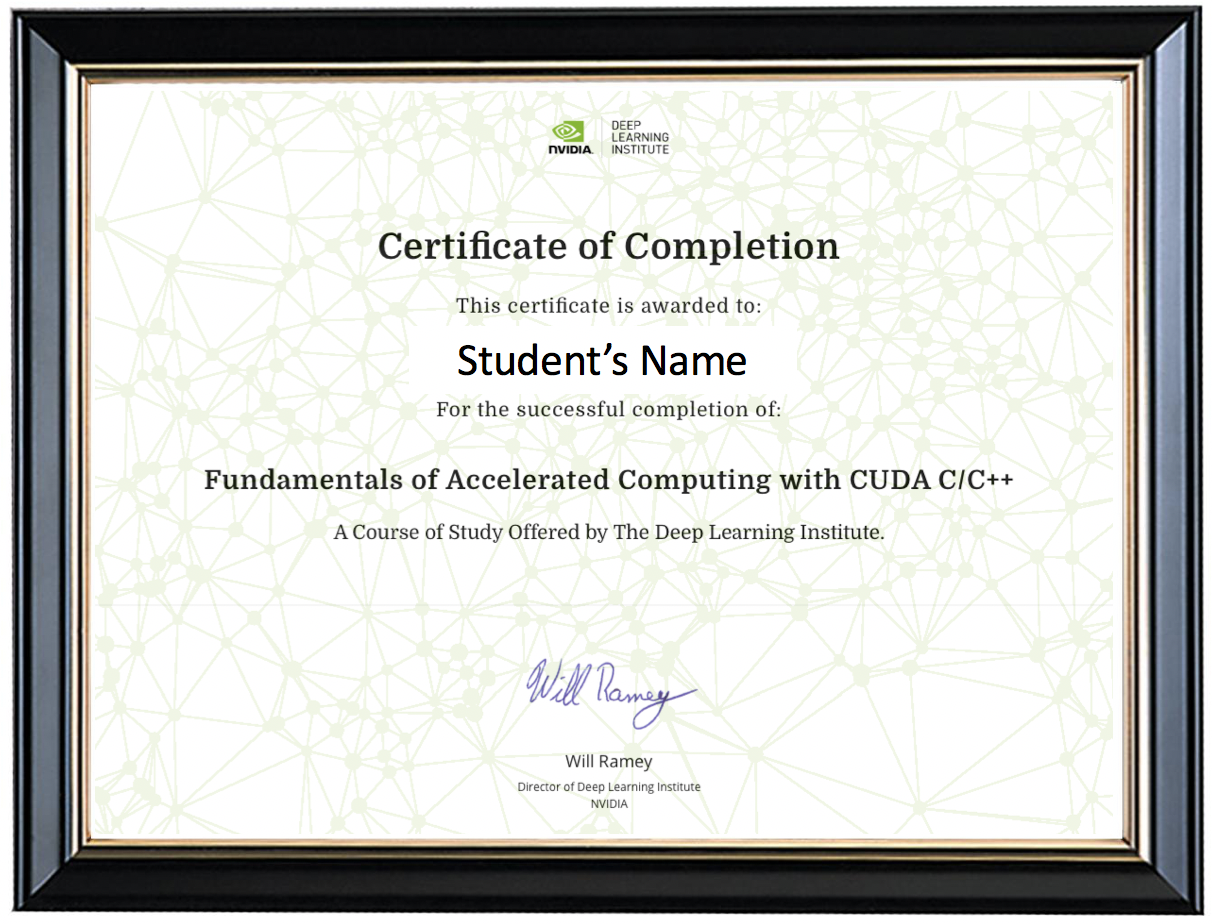}
\caption{Certificate of Competency. After completing the course and passing the project-based assessment, students receive a certificate of course competency.}
\end{figure}

\section{Results}
DLI has awarded over 150 certified University Ambassadors across the globe, and they have delivered over 150 instructor-led workshops in less than a year's time since 2017. Many academics have included DLI course materials in their own syllabi to prepare their students to solve real-world problems using AI and parallel computing. 

\subsection{Reaching Students Globally}
At UC Berkeley, Prof. Rekesha and Prof. Arias have modified the content of their courses to include a full life-cycle example from the DLI of how Deep Learning can be used, and to illustrate the relationship between Deep Learning and CPU/GPU-powered hardware. At National Tsing Hua University, Prof. Lee has integrated DLI course materials into teaching activities. His team, NVISION, won the 2016 NVIDIA Intelligent Robotics challenge. 
	
In India, Prof. Chickerur of KLE Technology University has conducted several DLI training events training thousands of students. The feedback from students continues to be positive and favorable, especially due to the fact that some courses offer certification they can use to bolster their professional career in the future.

\subsection{Feedback from University Ambassador Instructor}
DLI content primarily focuses on the hands-on application of deep learning and parallel computing. This creates a good complement to the theoretical approaches of traditional academic curricula. DLI courses enable students who are new to deep learning quickly jump-start their journey of solving real world problems. The content is not designed to provide a low-level explanation on the core components of deep learning such as mathematics and statistics. 

At Tokyo Institute of Technology, Prof. Gutmann (author of the paper) conducted a pilot workshop using the CUDA course and online platform described in this paper, and there were no reported issues of students using the online learning platform despite the use of various host machines and operating system types. At the start of each topic within the course Prof. Gutmann would give a very brief introduction to the concepts, then the students were able to go through the material and solve the exercises that were corresponding to each topic. Unlike traditional lectures where students have little interactivity with the content and tend to easily lose interest and not be actively engaged, Prof. Gutmann's students remained enthusiastic with the hands-on materials and teaching instructions. At the end of the course, many students were able to use what they learned from the course and their hands-on exercises to solve the assessment task of accelerating a particle system code, resulting in their individual certificate of competency. Students who did not finish on-sight were  able to finish over the next couple days on their own, as the course platform conveniently provides students perpetual access to the online version of the course.

At University of Kentucky (UK), students from the Association for Computing Machinery organization also conducted the DLI CUDA C/C++. DLI Certified University Ambassador Xi Chen (author of the paper) taught the workshop paving the way for students to have a better understanding of how to accelerate and parallelize applications using GPU computing. UK students were very excited about learning the concepts from the course materials, and also from each other, since Xi Chen himself happens to be a PhD student. The CUDA course demystified GPU programming for the students, and the hands-on exercises provide a platform that allows students to challenge each other with positive reinforcement. Many students expressed how the course made it simple to quickly learn how to harness the power of CUDA and GPUs, and they were interested in future DLI courses. 

\subsection{Feedback from Students}
The feedback from students who attended the DLI CUDA C/C++ course was quite positive and encouraging, shown in Figure 11. None of the students had difficulty navigating the course platform, and all of them expressed interest in future DLI workshops in AI and accelerated computing. The feedback survey showed that almost all of the students thought the instructors were helpful during the learning path, with 86\% of them indicating the instructors were extremely helpful. Part of this can be attributed to DLI's rigorous instructor certification process consisting of a collection of instructor assessments and in-person interviews with DLI master instructors. When asking how useful the information provided by the course might be for solving real-world problems and science, 71\% of the students suggested the information is extremely useful, while 29\% indicated moderately. Regarding the question about how helpful the course was for sharpening their HPC programming knowledge in general, 72\% suggested it was extremely helpful, 14\% expressed moderately helpful, and 14\% indicated slightly helpful.

\begin{figure}
\includegraphics[width=0.45\textwidth]{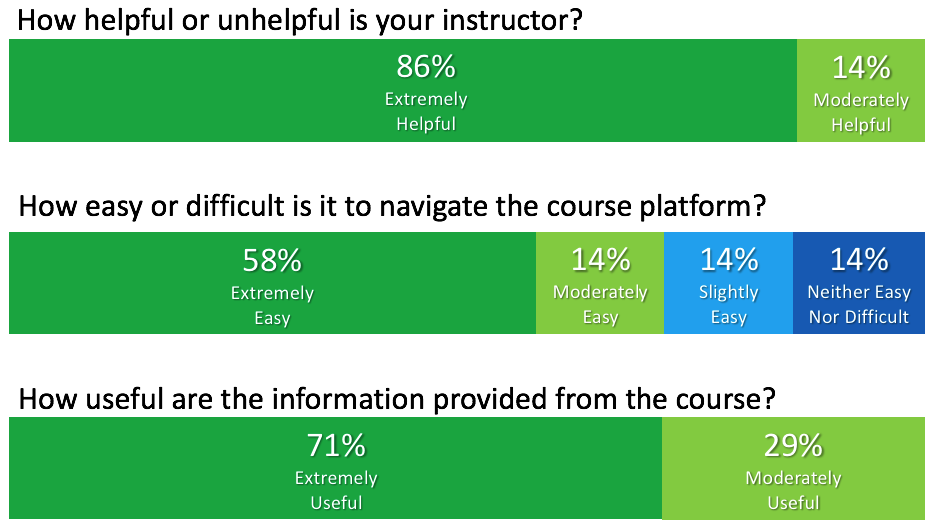}
\caption{Summary of the Student Feedback Survey for Fundamentals of Accelerated Computing with CUDA C/C++.}
\end{figure}

We also asked what percentage of the course should be instructor-led vs. self-paced, and students suggested they should be roughly 50\% versus 50\%. There were a few suggestions on how to improve the course delivery: 66\% expressed a desire to add more low-level details and theory in the course, and 17\% wish the course content was easier to understand. Other free-form comment responses from the survey included more time was needed for the exercises and more practical examples and less number crunching would be beneficial.

\section{Conclusions}
Like the discovery of the electricity, Deep Learning is fundamentally changing the world using scalable HPC platforms as the medium of education for this technology. Major breakthroughs in computer vision, Natural Language Processing (NLP) and autonomous driving go hand-in-hand with the growth of HPC. Three related goals form the foundation of DLI course design: (1) to provide training to assist the AI and HPC communities in fulfilling its educational needs; (2) to help researchers transition to using the technologies future HPC will build upon; and (3) to explore novel learning and application-based teaching paradigms. We present the platform, content and programs from NVIDIA's Deep Learning Institute, which aims to train researchers, scientists and professional engineers how to solve real-world problems with AI and accelerated computing. We believe that deeper hands-on experiences provide the most rapid and effective style of technology education.

\begin{acks}
    The authors would like to thank NVIDIA Deep Learning Institute for providing the figures, documents and online course platform, and Tokyo Institute of Technology and ACM University of Kentucky chapter for their helps organizing the course workshops.

  The authors would also like to thank the anonymous reviewers for their valuable comments and helpful suggestions.

\end{acks}

\bibliographystyle{ACM-Reference-Format}
\bibliography{sample-bibliography}


\begin{thebibliography}{11}


\ifx \showCODEN    \undefined \def \showCODEN     #1{\unskip}     \fi
\ifx \showDOI      \undefined \def \showDOI       #1{#1}\fi
\ifx \showISBNx    \undefined \def \showISBNx     #1{\unskip}     \fi
\ifx \showISBNxiii \undefined \def \showISBNxiii  #1{\unskip}     \fi
\ifx \showISSN     \undefined \def \showISSN      #1{\unskip}     \fi
\ifx \showLCCN     \undefined \def \showLCCN      #1{\unskip}     \fi
\ifx \shownote     \undefined \def \shownote      #1{#1}          \fi
\ifx \showarticletitle \undefined \def \showarticletitle #1{#1}   \fi
\ifx \showURL      \undefined \def \showURL       {\relax}        \fi
\providecommand\bibfield[2]{#2}
\providecommand\bibinfo[2]{#2}
\providecommand\natexlab[1]{#1}
\providecommand\showeprint[2][]{arXiv:#2}

\bibitem[\protect\citeauthoryear{Bernstein}{Bernstein}{2014}]%
        {Bernstein:2014dv}
\bibfield{author}{\bibinfo{person}{David Bernstein}.}
  \bibinfo{year}{2014}\natexlab{}.
\newblock \showarticletitle{{Containers and Cloud: From LXC to Docker to
  Kubernetes}}.
\newblock \bibinfo{journal}{\emph{IEEE Cloud Computing}} \bibinfo{volume}{1},
  \bibinfo{number}{3} (\bibinfo{date}{Sept.} \bibinfo{year}{2014}),
  \bibinfo{pages}{81--84}.
\newblock


\bibitem[\protect\citeauthoryear{Cook}{Cook}{2013}]%
        {Cook:2013cq}
\bibfield{author}{\bibinfo{person}{Shane Cook}.}
  \bibinfo{year}{2013}\natexlab{}.
\newblock \bibinfo{booktitle}{\emph{{CUDA Hardware Overview}}}.
\newblock \bibinfo{publisher}{Morgan Kaufmann}.
\newblock


\bibitem[\protect\citeauthoryear{Epstein, Lazarus, Calvano, Matthews, Hendel,
  Epstein, and Brosvic}{Epstein et~al\mbox{.}}{2017}]%
        {Epstein:2017ij}
\bibfield{author}{\bibinfo{person}{Michael~L Epstein}, \bibinfo{person}{Amber~D
  Lazarus}, \bibinfo{person}{Tammy~B Calvano}, \bibinfo{person}{Kelly~A
  Matthews}, \bibinfo{person}{Rachel~A Hendel}, \bibinfo{person}{Beth~B
  Epstein}, {and} \bibinfo{person}{Gary~M Brosvic}.}
  \bibinfo{year}{2017}\natexlab{}.
\newblock \showarticletitle{{Immediate Feedback Assessment Technique Promotes
  Learning and Corrects Inaccurate first Responses}}.
\newblock \bibinfo{journal}{\emph{The Psychological Record}}
  \bibinfo{volume}{52}, \bibinfo{number}{2} (\bibinfo{date}{May}
  \bibinfo{year}{2017}), \bibinfo{pages}{187--201}.
\newblock


\bibitem[\protect\citeauthoryear{Ho, Chuang, Reich, Coleman, Whitehill,
  Northcutt, Williams, Hansen, Lopez, and Petersen}{Ho et~al\mbox{.}}{2015}]%
        {Ho:2015ee}
\bibfield{author}{\bibinfo{person}{Andrew Ho}, \bibinfo{person}{Isaac Chuang},
  \bibinfo{person}{Justin Reich}, \bibinfo{person}{Cody Coleman},
  \bibinfo{person}{Jacob Whitehill}, \bibinfo{person}{Curtis Northcutt},
  \bibinfo{person}{Joseph Williams}, \bibinfo{person}{John Hansen},
  \bibinfo{person}{Glenn Lopez}, {and} \bibinfo{person}{Rebecca Petersen}.}
  \bibinfo{year}{2015}\natexlab{}.
\newblock \showarticletitle{{HarvardX and MITx: Two Years of Open Online
  Courses Fall 2012-Summer 2014}}.
\newblock \bibinfo{journal}{\emph{SSRN Electronic Journal}}
  (\bibinfo{date}{March} \bibinfo{year}{2015}).
\newblock


\bibitem[\protect\citeauthoryear{Luebke and {2008}}{Luebke and {2008}}{[n.
  d.]}]%
        {Luebke:ih}
\bibfield{author}{\bibinfo{person}{David Luebke} {and}
  \bibinfo{person}{{2008}}.} \bibinfo{year}{[n. d.]}\natexlab{}.
\newblock \showarticletitle{{CUDA: Scalable parallel programming for
  high-performance scientific computing}}. In \bibinfo{booktitle}{\emph{2008
  5th IEEE International Symposium on Biomedical Imaging (ISBI 2008)}}.
  \bibinfo{publisher}{IEEE}, \bibinfo{pages}{836--838}.
\newblock


\bibitem[\protect\citeauthoryear{Mullen, Byun, Gadepally, Samsi, Reuther, and
  Kepner}{Mullen et~al\mbox{.}}{2017}]%
        {Mullen:2017ih}
\bibfield{author}{\bibinfo{person}{Julia Mullen}, \bibinfo{person}{Chansup
  Byun}, \bibinfo{person}{Vijay Gadepally}, \bibinfo{person}{Siddharth Samsi},
  \bibinfo{person}{Albert Reuther}, {and} \bibinfo{person}{Jeremy Kepner}.}
  \bibinfo{year}{2017}\natexlab{}.
\newblock \showarticletitle{{Learning by doing, High Performance Computing
  education in the MOOC era}}.
\newblock \bibinfo{journal}{\emph{J. Parallel and Distrib. Comput.}}
  \bibinfo{volume}{105} (\bibinfo{date}{July} \bibinfo{year}{2017}),
  \bibinfo{pages}{105--115}.
\newblock


\bibitem[\protect\citeauthoryear{Nickolls, Buck, Garland, and Skadron}{Nickolls
  et~al\mbox{.}}{2008}]%
        {Nickolls:2008hn}
\bibfield{author}{\bibinfo{person}{John Nickolls}, \bibinfo{person}{Ian Buck},
  \bibinfo{person}{Michael Garland}, {and} \bibinfo{person}{Kevin Skadron}.}
  \bibinfo{year}{2008}\natexlab{}.
\newblock \showarticletitle{{Scalable parallel programming with CUDA}}. In
  \bibinfo{booktitle}{\emph{ACM SIGGRAPH 2008 classes}}.
  \bibinfo{publisher}{ACM Press}, \bibinfo{address}{New York, New York, USA},
  \bibinfo{pages}{1}.
\newblock


\bibitem[\protect\citeauthoryear{Porter, Haseltine, Conference, and
  {2015}}{Porter et~al\mbox{.}}{[n. d.]}]%
        {Porter:92vs5iLn}
\bibfield{author}{\bibinfo{person}{B Porter}, \bibinfo{person}{M Haseltine},
  \bibinfo{person}{N~Batchelder The Open~edX Conference}, {and}
  \bibinfo{person}{{2015}}.} \bibinfo{year}{[n. d.]}\natexlab{}.
\newblock \bibinfo{title}{{The state of Open edX}}.
\newblock


\bibitem[\protect\citeauthoryear{{Ravi, Daniele}, {Wong, Charence},
  {Deligianni, Fani}, {Berthelot, Melissa}, {Andreu-Perez, Javier}, {Lo,
  Benny}, and {Yang, Guang-Zhong}}{{Ravi, Daniele} et~al\mbox{.}}{[n. d.]}]%
        {Ravi:dw}
\bibfield{author}{\bibinfo{person}{{Ravi, Daniele}}, \bibinfo{person}{{Wong,
  Charence}}, \bibinfo{person}{{Deligianni, Fani}},
  \bibinfo{person}{{Berthelot, Melissa}}, \bibinfo{person}{{Andreu-Perez,
  Javier}}, \bibinfo{person}{{Lo, Benny}}, {and} \bibinfo{person}{{Yang,
  Guang-Zhong}}.} \bibinfo{year}{[n. d.]}\natexlab{}.
\newblock \showarticletitle{{Deep Learning for Health Informatics}}.
\newblock \bibinfo{journal}{\emph{IEEE journal of biomedical and health
  informatics}} \bibinfo{volume}{21}, \bibinfo{number}{1} (\bibinfo{year}{[n.
  d.]}), \bibinfo{pages}{4--21}.
\newblock


\bibitem[\protect\citeauthoryear{Silver, Huang, Maddison, Guez, Sifre, van~den
  Driessche, Schrittwieser, Antonoglou, Panneershelvam, Lanctot, Dieleman,
  Grewe, Nham, Kalchbrenner, Sutskever, Lillicrap, Leach, Kavukcuoglu, Graepel,
  and Hassabis}{Silver et~al\mbox{.}}{2016}]%
        {Silver:2016hl}
\bibfield{author}{\bibinfo{person}{David Silver}, \bibinfo{person}{Aja Huang},
  \bibinfo{person}{Chris~J Maddison}, \bibinfo{person}{Arthur Guez},
  \bibinfo{person}{Laurent Sifre}, \bibinfo{person}{George van~den Driessche},
  \bibinfo{person}{Julian Schrittwieser}, \bibinfo{person}{Ioannis Antonoglou},
  \bibinfo{person}{Veda Panneershelvam}, \bibinfo{person}{Marc Lanctot},
  \bibinfo{person}{Sander Dieleman}, \bibinfo{person}{Dominik Grewe},
  \bibinfo{person}{John Nham}, \bibinfo{person}{Nal Kalchbrenner},
  \bibinfo{person}{Ilya Sutskever}, \bibinfo{person}{Timothy Lillicrap},
  \bibinfo{person}{Madeleine Leach}, \bibinfo{person}{Koray Kavukcuoglu},
  \bibinfo{person}{Thore Graepel}, {and} \bibinfo{person}{Demis Hassabis}.}
  \bibinfo{year}{2016}\natexlab{}.
\newblock \showarticletitle{{Mastering the game of Go with deep neural networks
  and tree search}}.
\newblock \bibinfo{journal}{\emph{Nature}} \bibinfo{volume}{529},
  \bibinfo{number}{7587} (\bibinfo{date}{Jan.} \bibinfo{year}{2016}),
  \bibinfo{pages}{484--489}.
\newblock


\bibitem[\protect\citeauthoryear{Szegedy, Liu, Jia, Sermanet, and Reed}{Szegedy
  et~al\mbox{.}}{2015}]%
        {Szegedy:2015tb}
\bibfield{author}{\bibinfo{person}{C Szegedy}, \bibinfo{person}{W Liu},
  \bibinfo{person}{Y Jia}, \bibinfo{person}{P Sermanet}, {and}
  \bibinfo{person}{S Reed}.} \bibinfo{year}{2015}\natexlab{}.
\newblock \showarticletitle{{Going deeper with convolutions}}.
\newblock  (\bibinfo{year}{2015}).
\newblock


\end{thebibliography}

\end{document}